\documentclass[aps,prl,twocolumn,groupedaddress,floatfix,showpacs]{revtex4}
\usepackage{amsmath}
\usepackage{graphicx}
\usepackage{amsfonts}
\usepackage{amssymb}
\input{epsf}

\begin{document}

\title{Jamming, two-fluid behavior and `self-filtration' 
in concentrated particulate suspensions}
\author{M. D. Haw}
\email{M.Haw@ed.ac.uk}
\affiliation{School of Physics, University of Edinburgh, Kings Buildings,
Mayfield Rd., Edinburgh EH9 3JZ, U.K. \\}

\date{\today}
\begin{abstract}
We study the flow of model experimental hard sphere colloidal
suspensions at high volume fraction $\Phi$ 
driven through a constriction by a pressure gradient.  Above a particle-size
dependent limit
$\Phi_0$, direct microscopic observations
demonstrate jamming and unjamming---conversion of fluid to solid and vice versa---
during flow.  We show that such a jamming flow produces a reduction in 
colloid concentration $\Phi_{x}$ downstream of the constriction.
We propose that this `self-filtration' effect is the consequence of a combination 
of jamming of the \emph{particulate} part of the system and continuing flow of the
 \emph{liquid} part, 
i.e. the solvent, through the pores of the jammed solid.  Thus we link the 
concept of jamming in colloidal and granular media with a
`two-fluid'-like picture of the flow of concentrated suspensions.
Results are also discussed in the light of Osborne Reynolds' original experiments on
dilation in granular materials.
\end{abstract}



\maketitle

In this paper we consider the pressure-driven flow of concentrated
suspensions of model colloidal particles.
Concentrated suspensions of particles in liquid solvents are ubiquitous in
`soft matter' technology (cosmetics, foods, building
materials, paints, detergents, pharmaceuticals, waste management) as well as
in natural phenomena (soil and wet sand, formation of porous rocks and
sediments, landslip, etc). Much industrial processing 
of soft matter, and many examples of natural flow phenomena, involve
pressure-driven flow; moreover the flow often features `complex' geometries
where convergent and divergent elements
generate extensional components of strain (e.g. constricting and widening
pipes in a transport system). Fundamental studies of suspension
behavior in such complex flow remain rare in comparison to the wealth of
studies of soft matter under simple shear \cite{Larsonrheobook}.
Rheometrical work on concentrated suspensions 
has demonstrated complicated effects such as stress-induced thickening, erratic flow response,
and fluctuating viscosity \cite{Launerratic,
Friththickening}.

In the rheology of very concentrated suspensions
 and other crowded soft
matter systems a concept that has excited much recent speculation
\cite{NatureNV} and theoretical and experimental work
\cite{Bibettejamming,Colin,Veronique,Catesfragile,LiuNagelbook} is that of \emph{jamming}. 
Here we shall define jamming as
\emph{the conversion of a liquid system into a
solid by imposed stress}. Jamming is very obvious in `hourglass'
flow of dry sand where stress-supporting solid arches form across
the convergence, even though the sand typically flows more or less like a
liquid in simpler geometries. 
`Dilation' of wet sand is a related example with a venerable history,
having been considered more than 100 years ago by Reynolds \cite{Reynolds}.
However there remains no clear picture of the generic conditions required for nor the
consequences of jamming in soft matter. Our study of a model system is
aimed toward such a goal.

Our experimental system consists of polymethylmethacrylate (PMMA) spheres
sterically stabilised by short grafted polymers (polyhydroxystearyl alcohol),
suspended in a non-polar hydrocarbon solvent (decalin). This system is a good model 
of the conceptual
`hard sphere' suspension \cite{Puseyreview}. The particles can be
manufactured in almost monodisperse batches (polydispersities typically less
than 6-7\%).  Here we
study three batches of particles of radii $318 \pm 10$ nm, $656 \pm 20$ nm and 
$1000 \pm 50$ nm respectively, spanning the particle size range from
`colloidal'  (where Brownian motion dominates in dilute suspensions) toward `granular'
 where gravity is dominant.

We first demonstrate quantitatively
one important and striking consequence of jamming in a concentrated suspension:
that jamming in a convergent flow can induce a significant reduction in concentration
downstream of the constriction,
what one might call a `self-filtration' behavior by the suspension.
We study the effect
of convergent flow on concentration by extracting a portion of a bulk sample
through a constriction---put simply,
by sucking a small volume out of a bulk cell through a narrow-barrelled
syringe and comparing the resulting `extracted' concentration $\Phi_x$ with the bulk
concentration $\Phi_b$.

Samples are prepared as follows.  Dilute suspensions are
centrifuged until the particles form `solid' sediments (see below for a
discussion of jamming effects on sedimentation). To obtain given
bulk volume fractions measured amounts of solvent are added to the
sediments.  In practice we prepare
\emph{pairs} of samples which have been subjected to exactly the same 
treatment (i.e. duration and
rate of centrifugation etc.\ ) and which thus have equal $\Phi_b$ to within the
(small) uncertainties associated with measuring masses of sediment and added solvent.   
(We use \emph{pairs} of samples for the comparison because, if $\Phi_x \ne \Phi_b$, extracting 
some sample by syringe to measure $\Phi_x$ and then measuring $\Phi_b$ for the \emph{same}
sample would introduce systematic errors.)
For one member of the pair we obtain
the bulk colloidal \emph{mass} fraction by scooping out a measured mass of the
sample, allowing all solvent to evaporate, and measuring the mass
of solids remaining.  
From the other member of the pair we extract a portion not by scooping but by \emph{syringe},
and similarly measure the mass fraction of this extracted portion.
We convert mass fractions
into volume fractions $\Phi$ (the more common measure of concentration in studies of
colloidal suspensions) assuming a particle density $\rho_p = 1.188$ g cm$^{-3}$ (the
bulk density of PMMA) and solvent
density $\rho_s = 0.897$ g cm$^{-3}$.
If there is no effect of convergent flow in the syringe on
extracted volume fraction we should find $\Phi_x = \Phi_b$.  A large
number of experiments
is carried out on rediluted and newly centrifuged pairs of samples to ensure that 
statistical variations due to measurement imprecision are minimised.  


There are a number of ways of estimating $\Phi$ for a colloidal suspension, none of
which is without drawbacks. As is well known
\cite{PuseyHScrystals} the `mass fraction' measure of $\Phi$ tends to give systematically
lower values compared to the other common measure in the PMMA system, that based on 
mapping experimental phase
boundaries onto computer simulated fluid-crystal hard-sphere phase coexistence
boundaries. Given the core-shell nature of the sterically-stabilised
particles no simple measure can give an \emph{absolutely} correct result
in the sense of a `true hard sphere' volume fraction. The important point for
this work is that all measures are carried out in the same way so that we obtain volume
fractions that are consistent and comparable with each other. In these
experiments, given the small volumes (see below) of extracted suspension, 
the possible experimental errors are minimised by
using the `mass fraction' to obtain $\Phi$; dilution and mapping onto phase boundaries
is simply not possible with such small sample volumes.  
However care must be taken when
comparing the numerical values of $\Phi$ given here with those quoted elsewhere in
the literature, which are often obtained by mapping onto the computer-simulation
phase diagram.

%
%
%
%

%
%
%
%

All experiments are carried out in identical geometry: the `bulk' cells are
cylindrical glass cuvettes of diameter $25$ mm with sediment heights
$\approx 2.5$ cm giving a total sample volume of $\approx 12$ cm$^{3}$; the
syringes used have maximum volume $1.0$ ml with an entry barrel internal
diameter of $1.6$ mm;
volumes extracted are typically $0.4-0.5$ ml, i.e. always $< 5\%$ of the
total bulk volume.  The `sucking' 
procedure is as follows: the syringe plunger is pulled out quickly 
by hand to generate an empty barrel volume of $\approx 0.6$ ml, so that
a pressure drop ($P_b/P_{atm} \approx 0.03$) 
is instantaneously applied between
the outside surface of the bulk at atmospheric pressure $P_{atm}$ and the surface inside
the syringe barrel at reduced pressure $P_b$.  
In the majority of the experiments reported here we study this limiting case of 
a suddenly applied initial pressure drop (the pressure drop
slowly decreases as
the sample is extracted and enters the syringe barrel). 
A few experiments have also been carried out
using larger initial pressure drop and using narrower constrictions.

\begin{figure}
\begin{center}
  \epsfysize=6cm
   \leavevmode\epsffile{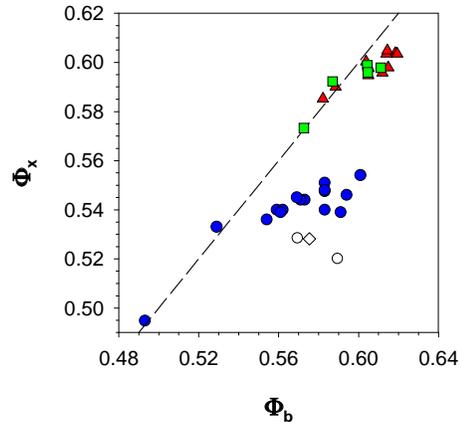}
	\caption{Bulk volume fraction $\Phi_b$ \emph{vs} volume fraction extracted by
   syringe $\Phi_x$.  The line indicates $\Phi_b = \Phi_x$, i.e. the expected result
   if convergent flow into syringe has no `self-filtration' effect.  Filled circles, 
 empty diamond and empty circles, particle
 radius 1000nm; filled triangles, radius 656nm; filled squares, 318nm.  Empty circles show
experiments extracting through a fine needle rather than syringe barrel; the diamond
shows an experiment using a larger initial pressure drop.  Error bars on $\Phi$ are 
$\sim \pm 0.005$, i.e. approximately the size of the symbols.}
\label{fig_graph}
\end{center}
\end{figure}

Results for $\Phi_{b}$ \emph{vs} $\Phi_{x}$ are shown in
Figure~\ref{fig_graph}, for the three particle sizes given above.  As is clear,
when $\Phi_{b}$ is above some limit volume fraction $\Phi_{0}$ 
the sample 'pumped' through the contraction into the syringe barrel has a 
significantly reduced
volume fraction $\Phi_{x}<\Phi_{b}$.  The limit concentration $\Phi_{0}$ is
strongly dependent on particle size, decreasing with increasing particle size.
Thus the effect is more visible with larger particles (approaching the granular
scale) but even for the smallest particles, very much within the
\emph{colloidal} regime, there is a clear effect at the highest volume
fractions.  The reduction in extracted
volume fraction is
somewhat more severe with a smaller constriction geometry or with a larger initial
pressure drop (see example open circles and
open diamond, respectively, in Fig.~\ref{fig_graph}), though we have yet to
carry out detailed investigations on the effect of different pressure drop. 
In any case, since extraction of
samples from bulk for e.g. observation in a microscope or measurement in
a rheometer is very often achieved using syringes, pipettes etc., it is
important for experimentalists to be aware of the sensitivity of key parameters
such as volume fraction to such `processing' prior to experiments.
Of course similar flow situations involving concentrated suspensions are common
in technological applications such as product delivery, waste processing, and
so on.


Although it seems that this `self-filtration' effect has not been
reported in the colloidal literature, actually it is not
too surprising if we consider the suspension as a `two-fluid' system, that is 
a combination of \emph{particulate fluid} and
\emph{liquid solvent}, where interaction between the particles may lead 
to a rheological `separation' of the two fluids.  There are a few studies of
\emph{squeeze flow} where a separation between solvent and dispersed phase is
discussed \cite{Lequeuxsqueeze,Delhayesqueeze}, but such a two-fluid picture of
flow, though familiar in studies of polymer systems, is not often encountered in the 
colloidal literature.  We propose that our results
can be explained in a two-fluid spirit as an `extreme' rheological separation of
the particulate and solvent parts of the suspension.  
The particles \emph{jam} and form a solid that resists the pressure drop and does not flow at 
all.  Meanwhile the solvent, remaining liquid, cannot
resist the pressure gradient and continues to flow through the pores of the jammed colloid.  
Hence there
is an increased flow rate of solvent relative to particles, resulting in a
reduced downstream particle volume fraction.

\begin{figure}
\begin{center}
  \epsfxsize=7cm
   \leavevmode\epsffile{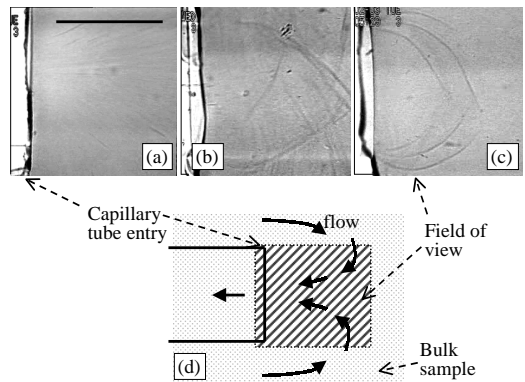}
	\caption{Images of the flow of a suspension of 1000 nm colloids, volume
fractions (a) $\Phi=0.534$; (b) $\Phi=0.578$; (c) $\Phi=0.60$.  The horizontal bar in (a)
is 0.5 mm.  At the left extreme of
each image the 
entrance to the capillary can be seen, as indicated by the schematic in (d) showing the 
viewing and flow geometry. 
In (b) and (c) dark lines form in arch-like 
patterns around the end of the capillary, while flow in (a) is smooth.  Still-images are taken
from digitised movies, see Ref.~\protect{\cite{jamming_WWW}} experiments 2, 3 and 6 for (a), (b),
 and (c) respectively.}
\label{fig_phicomp}
\end{center}
\end{figure}

\begin{figure}
\begin{center}
  \epsfxsize=7cm
   \leavevmode\epsffile{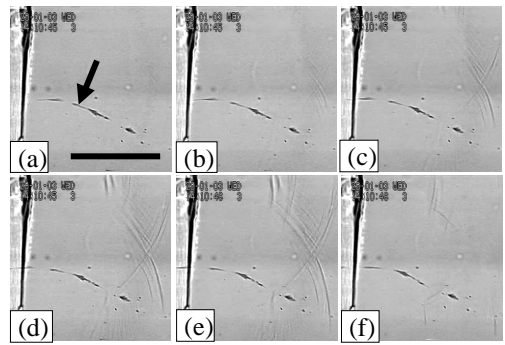}
	\caption{A sequence of images of the flow of a suspension of $318$ nm colloids, volume
fraction $\phi=0.61$.  The stills (a) to (f) are each separated by 
0.08 seconds.  At the left extreme of each image the 
entrance of the capillary can be seen.  In the top left image, the horizontal bar is $0.5$ mm,
while the arrow shows a 
line of dust particles from whose motion and distortion one may obtain an approximate idea of 
the velocity field flowing into the capillary entrance.  Dark lines form [toward the right hand
 side of images (b) to (e)] 
in arch-like  crossing configurations, then disappear very suddenly in (f) as the jam
`collapses'.  See movie at Ref.~\protect{\cite{jamming_WWW}} experiment~1.}
\label{fig_seq1}
\end{center}
\end{figure}

But is colloidal jamming really responsible for the measurements in Fig~\ref{fig_graph}?  
To elucidate directly what is actually happening in our convergent pressure-driven
flow we have carried out direct observations by optical microscope.  
To enable microscope observations, 
the PMMA particles are suspended in a mixture of solvents (decalin and
tetralin) to partly match the particle refractive index and reduce
multiple scattering, and samples are prepared
(centrifuged) inside thin \emph{rectangular} cuvettes.  
`Extraction' is achieved by inserting
a cylindrical glass capillary into the sediment, internal diameter $1.0$ mm, 
connected to a syringe whose plunger is
withdrawn using a fixed-speed syringe pump.  The 
flow in the region of the capillary tube entrance is observed with bright
field microscopy, the field of view positioned at the region around the
entrance to the capillary as shown in Fig.~\ref{fig_phicomp}(d).  
Results in the form of digitised movies 
are available at \cite{jamming_WWW}, while figures~\ref{fig_phicomp}(a)-(c) 
and \ref{fig_seq1} show extracted
still-images.
In all but one of the movies, the height of the field of view 
corresponds to $\sim 1$ mm and the entry to the capillary can be seen at
the left of the picture.  At such low magnification individual colloids cannot be
resolved.  In one movie (Ref.~\cite{jamming_WWW} experiment~4) the magnification is
 increased by a factor of 10 and
the field of view is centred $\sim 0.5$ mm diagonally from the lower corner of the
capillary entry; in this case the particles, $1000$ nm radius, are just resolvable.


Summarising our observations, flows at high $\Phi$ are typically very erratic, 
demonstrating very clear \emph{transient jamming} of the samples in
the region of the tube entrance, involving slowing of flow, sudden `fracture' events 
followed by speed-up, repeated re-jamming, and so on.  
Jamming as a transient `conversion' of flowing liquid to stationary 
solid is strikingly apparent.  In the movie at higher magnification 
(Ref.~\cite{jamming_WWW} experiment~4) the sample, blurred whilst flowing (i.e. liquid), 
can be seen to repeatedly momentarily freeze (become solid).

At the highest concentrations flow is 
localised near the entry of the capillary (Ref.~\cite{jamming_WWW} experiment~6). 
However
dark lines form in arch-like shapes, often travelling in sudden `waves' or `shocks' 
out from
the end of the capillary [Figs.~\ref{fig_phicomp}(b) and (c) 
and Fig~\ref{fig_seq1}].  The `shocks' are associated with subsequent speed-up of the flow, 
i.e. collapses of the jammed particle structure.
Collapse is followed by rejamming, generating erratic changes in flow speed.  
The most likely optical origin of the dark lines
is small-angle scattering (their appearance is very sensitive to
refractive index contrast consistent with this), indicating localised changes in volume 
fraction.  
The exact nature of these waves or shocks of dark lines is unclear.  They
exhibit similarities with density or kinematic waves observed in granular media, such as
`hourglass' flow of sand \cite{Baxter_granular}.  
More detailed study and comparison with
granular density waves will be the subject of future work \cite{mewavestbp}.

As $\Phi_b$ is lowered toward the limit $\Phi_0$, the jam-collapse-rejam behavior cycles
faster (Ref.~\cite{jamming_WWW} experiments 3 and 4),
 suggesting that below some limiting concentration jams effectively collapse
immediately, in other words the sample no longer jams during flow.  Below $\Phi_0$,
 we indeed observe
smooth non-jamming flow of the suspension into the convergence [Fig.~\ref{fig_phicomp}(a) 
and Ref.~\cite{jamming_WWW} experiment~2].

Continued flow of the reduced volume fraction (partially `self-filtered')
sample downstream presumably accounts for the \emph{transience} of the jamming: 
jams give way
just as, in hourglass flow of sand, solid arches give way to allow flow.

Collapse of jams apparently involves a sliding-solid fracture-like behavior 
(Ref.~\cite{jamming_WWW} experiment~5),
sliding occurring at the dark lines visible in the sample.
We speculate that such sliding-fracturing may also be a consequence of
the combination of jamming and solvent permeation through the jam.  
Fractures may be localised by geometry-dependent pressure gradients
driving solvent through the pores of the jams into regions of decreased
particle concentration---sometimes called `microcracks' in the rock fracture
literature---which we associate with the dark lines 
we observe in the suspensions.
The excess solvent in these regions lubricates localised sliding (`cataclastic 
shear bands' \cite{Mainslidingrocks}) of
opposing solid regions.  Such a `lubricated slide' picture has
been proposed for slip of porous rocks in earthquakes (see e.g. \cite{Mainslidingrocks}).
The possibility that model 
suspensions might ultimately be used as simple `soft' analogues of geological systems
deserves further investigation.

A further demonstration of jamming is provided by simple observations of the
 behavior of sediments
of the largest particles. Though apparently solid under careful handling, we
have observed that the
sediment formed after centrifugation very easily \emph{reliquifies} under a
slight deliberate \emph{lateral} shaking. 
The sediment thus behaves as an example of what has
been called `fragile matter' \cite{Catesfragile}: it solidifies (jams) under
application of a \emph{unidirectional} force in the centrifuge (or over a longer
time period in normal
gravity) but cannot support (liquifies under) small stresses applied in
\emph{any other direction}.

Finally we note that the `sucking' experiments may be compared to the
early granular experiments of Osborne Reynolds, who squeezed rubber balloons 
filled with ball
bearings and water \cite{Reynolds}.  Reynolds observed that on applying pressure, water was 
drawn \emph{into} 
the balloon---quite the reverse of the case of a balloon completely 
filled with a simple fluid, which would be forced \emph{out}.  The ball-bearing
system inside the balloon can only strain by decreasing its local volume fraction, 
i.e. by \emph{dilation} 
and increase of the total volume of the flexible balloon.  Water is then naturally
drawn in to fill the extra volume.  Our observations in colloidal suspensions 
may also be interpreted in terms of dilation in the zone
of convergent flow as follows.  To allow flow in response to the applied pressure drop 
the region of sample just downstream of the convergence
must dilate, i.e. decrease its volume fraction by taking on more solvent; 
but this solvent must come from somewhere, i.e.
from the region upstream, hence generating an increase in upstream $\Phi$.  This
increase in $\Phi$ in
turn leads to \emph{jamming} of the colloidal particles.  


The author acknowledges the support of the Royal Society of Edinburgh and
thanks A. B. Schofield for the manufacture of the PMMA particles.

\vspace{-5mm}

\bibliography{jam}

%
\end{document}